# AUTOMATIC ANNOTATION OF XHTML PAGES WITH AUDIO COMPONENTS

**ABSTRACT**

In this paper we present Deiush, a multimodal system for browsing hypertext Web documents. The Deiush system is based on our novel approach to automatically annotate hypertext Web documents (i.e. XHTML pages) with browsable audio components. It combines two key technologies: (1) middleware automatic separation of Web documents through structural and semantic analysis which is annotated with audio components, transforming them into XHTML+VoiceXML format to represent multimodal dialog; and (2) Opera Browser, an already standardized browser which we adopt as an interface of the XHTML+VoiceXML output of annotating. This paper describes the annotation technology of Deiush and presents an initial system evaluation.

## 1. INTRODUCTION

In the past ten years significant research effort was invested in audio browsers, programs able to decode the structure of Web pages and put them into an audio format. Few advanced browsers use machine learning algorithms to classify objects on the Web page and learn browsing behaviors, have multimodal input and outputs and are able to synchronize between the graphical and audio modalities to interact with the Web page. The disadvantages of these audio browsers include the necessity of high computation power since the user's machine has to decode the structure of the Web page, and, therefore, making impossible the installation of such programs on mobile devices.

Most audio browsers existent today (e.g. GW Micro Window Eyes [GW Micro, 2006], IBM Home Page Reader [IBM, 2003], Freedom Scientific Jaws [Freedom Scientific, 2006], the systems from the W3C Alternative Web Browsing initiative [W3C, 2005]) are sequential screen readers. Few advanced browsers use multimodal input and outputs, or are able to synchronize between the graphical and audio modalities to interact with the Web page. Based on a simpler and more efficient solution for the creation of multimodal applications, we developed the multimodal (graphic and audio) system *Deiush*. Our design consists of a middleware that automatically annotates Web pages with VoiceXML generated from the content of the Web page. Using our system, the user can interact with the Web page using voice or graphical commands, can listen and/or watch digital content, such as news, blogs Web pages, podcasts, and even access emails and personal schedules.

In this paper, we present our approach to multimodal automatic annotation to Web pages and evaluate the system based on interaction speed and the user's experience.

## 2. METHODS

The structure of our middleware is depicted in Figure 1. The client connects to a middleware Web site and provides the Web page that he/she wants to access. The middleware application server collects the Web page, identifies the components in the HTML code and annotates them with VoiceXML (see the example in Figure 2a and b). The system synchronizes the XHTML code with VoiceXML components using "sync components". In this way, the user can interact with the Web page in two different ways: using the graphical components in the Web page or using the voice tools existent in the Opera browser. Since this interaction is synchronized, the user can interleave the modalities.

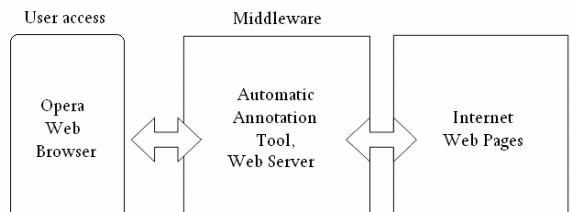

Figure 1. Automatic Annotation Architecture

```html
<select name="participants" id="participants" multiple="multiple" size="10" width="100%">
  <option> Anton, Tudor </option>
  <option> Cesar, Brian </option>
  <option> Danniels, David </option>
  <option> Tejada, Jose </option>
</select>
```

a. An HTML example

```xml
<?xml version="1.0" encoding="iso-8859-1"?>
<!DOCTYPE html PUBLIC "-//VoiceXML Forum//DTD XHTML+Voice 1.2//EN"
   "http://www.voicexml.org/specs/multimodal/x+v/12/dtd/xhtml+voice12.dtd">

<html xmlns="http://www.w3.org/1999/xhtml"
   xmlns:vxml="http://www.w3.org/2001/vxml"
   xmlns:ev="http://www.w3.org/2001/xml-events"
   xmlns:xv="http://www.voicexml.org/2002/xhtml+voice"
   xml:lang="en-US">

  <head>
    ...
    <vxml:form id="scheduler_meeting_form">
      ...
      <vxml:field name="voice_participants_name" xv:id="voice_participants_name" modal="true">
        <vxml:grammar>
          <![CDATA[
            #JSGF V1.0 iso-8859-1;
            grammar participants;
            public <participants> = <NULL> ( $= new Array; ) (<participant> [and] ( $.push($participant) ) )+;
            <participant> =  Anton, Tudor ($="Anton, Tudor") |
                             Cesar, Brian ($="Cesar, Brian") |
                             Danniels, David ($="Danniels, David") |
                             Tejada, Jose ($="Tejada, Jose") |
                             ... ;
          ]]>
        </vxml:grammar>
        <vxml:prompt bargein="true">
          Please say the participants.
        </vxml:prompt>
        <vxml:catch event="noinput">
          Sorry, I did not hear you. Please say the participants.
        </vxml:catch>
        <vxml:catch event="nomatch">
          Sorry, I did not understand you. Please say the participants.
        </vxml:catch>
      </vxml:field>
      ...
    </vxml:form>
    ...
    <xv:sync xv:field="#voice_participants_name" xv:input="participants"/>
    ...
  </head>
```

b. XHTML+VoiceXML result

Figure 2. VoiceXML Annotation: a. An HTML example, and b. The XHTML+VoiceXML result annotation

The components of the Web pages (i.e. the inputs, the outputs and the links to the following pages) are extracted from the Web page source and a tree is created with all possible dialog paths. The words are extracted from the collected HTML and are searched in a list of possible shortcuts and a ranking weight can be assigned to certain entries based on these words. The user can search for a certain subset of entries in the Web page by only pronouncing shortcut words. The grammar is automatically generated from the entries in the Web page. Additional VoiceXML units are added for verification purposes (see Figure 3).

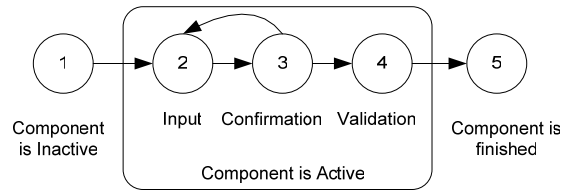

Figure 3. Verification audio component

## 3. RESULTS

A series experiments have been conducted to evaluate the accuracy and the quantitative performance of the Deiush against GW Micro Window Eyes, the state-of-the-art screen reader. In our evaluation we used our system on a simple case in which the user was dealing with transformed news and blog Web pages into XHTML+VoiceXML Web pages, and a more complex dialog for consulting a scheduler (see Figure 4). In all experiments, the input was achieved only through the audio input. In the complex application, Web developers used dynamic HTML, Javascript DOM and AJAX to update the results on the same Web page. However, since the generated voice components were synchronized with the HTML inputs, the result was the same for the dynamic output component.

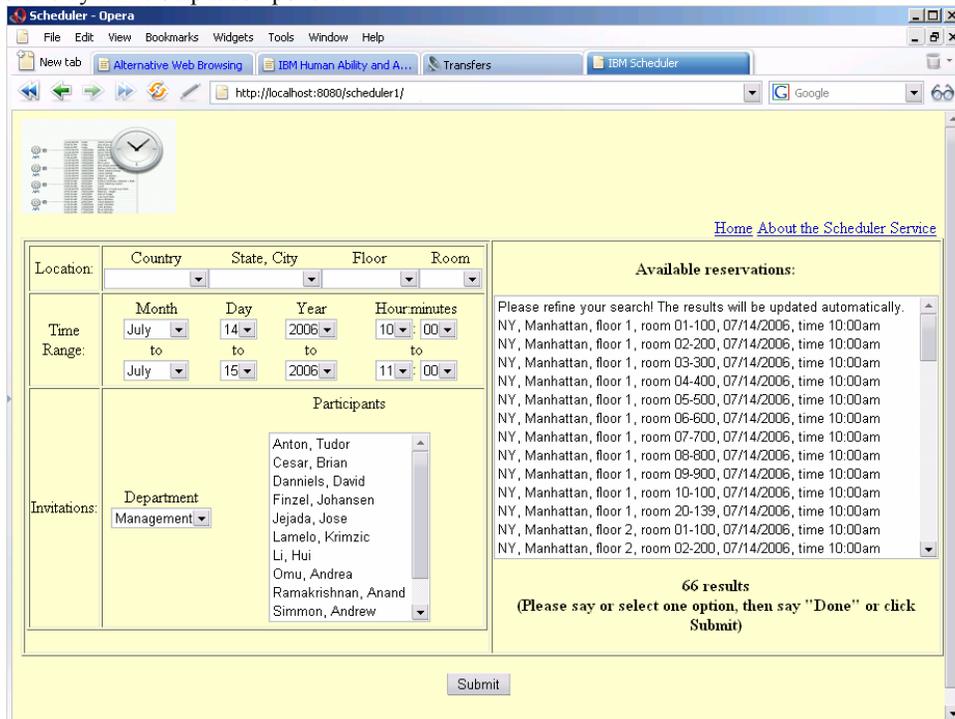

Figure 4. Multimodal Scheduler in Opera Browser

We used two metrics to compare our system with Window Eyes. First, we measured the average time to select the desired information and to complete one successful atomic step of interaction (considering the task completion success rate as 100%) (Figure 5). Second, we measured the quality cost, as the users' perception of the system. The user's satisfaction rating was computed by having users complete a survey at the end of the experiment (see Figure 6). In all the experiments we obtained results similar to the standard audio browser. The evaluation demonstrated that using a middleware does not require high user machine's computation power (to decode the structure of the Web page) and kept the same browsing experience for visually disabled individuals.

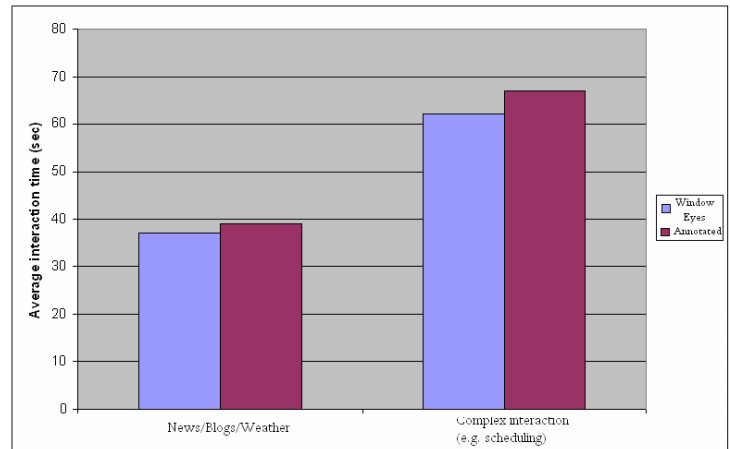

Figure 5. Average time to select the desired information from one page or to complete one successful atomic step of interaction

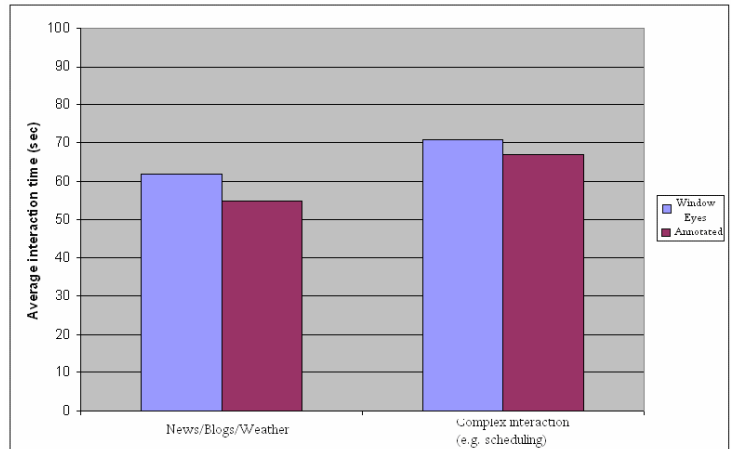

Figure 6. User satisfaction

## 4. CONCLUSION

In this paper, we have described the design and implementation of Deiush, a middleware-directed non-visual Web browsing system. We presented experimental evidence of the efficiency of middleware-directed Web browsing using audio. The Deiush system provided results similar to those of the audio browser, but did not require high computational power and additional software on the client device. The following are only a few potentially useful areas for further research. Machine learning techniques can be employed to rank and personalize dialog paths. Our text to speech generation unit can also be ported in the middleware, the client being a pure telephone device.

## REFERENCES

GW Micro, 2006, *Window Eyes*, http://www.gwmicro.com .

IBM, 2003, *IBM Home Page Reader*, http://www-03.ibm.com/able/ .

Freedom Scientific, 2006, *Jaws*, http://www.freedomscientific.com .

W3C, 2005, *Alternative Web Browsing initiative*, http://www.w3.org/WAI/References/Browsing .